# R.F. Pollution Reduction in Cellular Communication

Sumit Katiyar , Prof. R. K. Jain, Prof. N. K. Agrawal

**Abstract**— R. F. pollution has been recognized as health hazard in India in the prevailing circumstances. There is lot of hue and cry against cellular towers installed in residential area. Recently high court in India has issued an order not to install towers in residential areas. For meeting the exponential demand of cellular communication in India this will be a set back for future growth. An appropriate solution has to be developed for meeting demand as well as RF pollution concern of the society. This paper deals with the installation of low power base stations in residential areas instead of high power macro cell base stations. Macro stations are proposed to be used for fast traffic, low power micro cell for a slow traffic / pedestrian and pico cell / femto cell for indoor use. These cells will be in hierarchical structure along with adaptive frequency allocation techniques and A-SDMA approach.

——————————— ◆ ———————————

## 1 INTRODUCTION

RF pollution has been considered as health hazard for man kind. There were two opinions in this matter. As per first opinion, there is no scientific evidence for harmful effects of RF pollution at these RF levels which are being spread in the environment by cellular base stations. Lot of research has been conducted in this connection by different concerned departments and harmless levels were also predicted. As per other opinion, R.F. radiation is harmful to mankind and can cause neurological, cardiac, respiratory, ophthalmological, dermatological and other conditions ranging in severity from headaches, fatigue and add to pneumonia, psychosis and strokes. In India we didn't bother for R.F. pollution and developed our cellular network in residential area also as per system requirement completely ignoring this aspect. Recently this issue has been raised by media, social activists and medical experts loudly. As a result high court issued an order to stop installation of cellular tower in residential area in U.P. (India).
In view of above, it is the need of hour to develop a cellular network which may not raise the R.F. level beyond harmless levels. At the same time this network should meet the increased demand raised due to exponential growth of cellular communication. Obviously solution lies with the use of low power transmitters in the network. This paper explains the ways and means for enhancement of spectral density and reduction of power consumption as well as R.F. pollution too.
Hierarchical structure could be the answer of this problem if it is developed with the help of low power transmitters. This paper presents the four layers hierarchical cellular network consists of macro cell for fast traffic (less than 10 W), micro cells (less than 1W) for slow traffic / pedestrians and pico cell (less than 200 mW) / femto cell (less than 100 mW) for indoor areas. Adaptive frequency allocation and A-SDMA approach has been incorporated in the system to taking care of optimum resource utilization.

Instead of proper cell planning, vendors in India are using high power micro cells for penetrating signal inside the building which is consuming more power and increasing level of pollution. In this paper, we have suggested simple technologies with proper planning of networking for meeting multiple requirements.

## 2 EFFECT OF R.F. POLLUTION ON HUMAN HEALTH

Studies have shown that human beings are bio electrical systems. The heart and the brain are regulated by internal bio-electrical signals. Environmental exposures to EMF can interact with fundamental biological processes in the human body and in some cases this may cause discomfort

1. Modulation signals are one important component in the delivery of EMF signals to which cells, tissues, organs and individuals can respond biologically. Modulating signals have a specific beat defined by how the signal varies periodically over time. Modulation signals may interfere with normal, nonlinear biological functions.
2. There have been growing public concern of possible adverse health effects due to EMF Radiation. The area of concern is the radiation emitted by the fixed infrastructure used in mobile telephony such as base stations and their antennas, which provide the link to and from mobile phones.
3. There are two distinct possibilities by which the Radio Frequency Radiation (RFR) exposure may cause biological effects. There are thermal effects caused by holding mobile phones close to the body. Secondly, there could be possible non-thermal effects from both phones and base stations.

a) Thermal Effects:- One effect of microwave radiation is dielectric heating, in which any dielectric material, (such as living tissue) is heated by rotation of polar molecules induced by the electromagnetic field. The thermal effect has

————————————————

- Sumit Katiyar is currently pursuing PhD degree program in electronics engineering from Singhania University, India, E-mail: sumitkatiyar@gmail.com
- Prof. R. K. Jain is currently pursuing PhD degree program in electronics engineering from Singhania University, India, E-mail: rkjain_iti@rediffmail.com
- Prof. N. K. Agrawal is a senior member of IEEE and life member of ISTE and ISCEE. E-mail: agrawalnawal@gmail.com

been largely referred to the heat that is generated due to absorption of EMF radiation. In the case of a person using a cell phone, most of the heating effect occurs at the surface of the head, causing its temperature to increase by a fraction of a degree. The brain blood circulation is capable of disposing the excess heat by increasing the local blood flow. However, the cornea of the eye does not have this temperature regulation mechanism. The Thermal effect leads to increase in body temperature.

b) Non-Thermal Effects:- The communication protocols used by mobile phone often result low frequency pulsing of the career signal. The non-thermal effect is reinterpreted as the normal cellular response to an increase in temperature. The Non-thermal effects are attributed to the induced electromagnetic effects inside the biological cells of the body which is possibly more harmful. People who are chronically exposed to low level wireless antenna emissions and users of mobile handsets have reported feeling several unspecific symptoms during and after its use, ranging from burning and tingling sensation in the skin of the head, fatigue, sleep disturbance, dizziness, lack of concentration, ringing in the ears, reaction time, loss of memory, headache, disturbance in digestive system and heart palpitation etc. There are reports indicating adverse health effects of cell phones which emit electro-magnetic radiation, with a maximum value of 50% of their energy being deposited when held close to the head.

4. Member Scientist, ICMR has indicated that the hot tropical climate of the country, low body mass index (BMI), low fat content of an average Indian as compared to European countries and high environmental concentration of radio frequency radiation may place Indians under risk of radio frequency radiation adverse effect [1].

## 3 EMF EXPOSURE LIMITS FROM MOBILE BASE STATIONS

Some countries in the world have specified for their own radiation level keeping in view the environmental and physiological factors. In order to protect the population living around base stations and users of mobile handsets, established new, low intensity based exposure standards. The new exposure guidelines are hundreds or thousands of times lower than those of Institute of Electrical & Electronics Engineers (IEEE), USA and ICNIRP. The exposure limit for RF field of some of the countries including countries that have lowered their limit, in cell phone frequency range of 900 MHz and 1800 MHz for example are as given under [1] :-

Table – 1: International Exposure Standards

| International Exposure limits for RF fields (1800 MHz) | |
|---|---|
| 12 W/m² | USA, Canada and Japan |
| 9.2 W/m² | ICNIRP and EU recommendation 1998 – Adopted in India |
| 9 W/m² | Exposure limit in Australia |
| 2.4 W/m² | Exposure limit in Belgium |
| 1.0 W/m² | Exposure limit in Italy, Israel |
| 0.5 W/m² | Exposure limit in Auckland, New Zealand |
| 0.45 W/m² | Exposure limit in Luxembourg |
| 0.4 W/m² | Exposure limit in China |
| 0.2 W/m² | Exposure limit in Russia (since 1970), Bulgaria |
| 0.1 W/m² | Exposure limit in Poland, Paris, Hungary |
| 0.1 W/m² | Exposure limit in Italy in sensitive areas |
| 0.095 W/m² | Exposure limit in Switzerland, Italy |
| 0.09 W/m² | ECOLOG 1998 (Germany) Precaution recommendation only |
| 0.001 W/m² | Exposure limit in Austria |

Following are the reference levels based on international and national guidelines [1].

Table – 2: Reference levels for the general public at 900 & 1800 MHz

| International Exposure limits for RF fields (1800 MHz) | |
|---|---|
| 12 W/m² | USA, Canada and Japan |
| 9.2 W/m² | ICNIRP and EU recommendation 1998 – Adopted in India |
| 9 W/m² | Exposure limit in Australia |
| 2.4 W/m² | Exposure limit in Belgium |
| 1.0 W/m² | Exposure limit in Italy, Israel |
| 0.5 W/m² | Exposure limit in Auckland, New Zealand |
| 0.45 W/m² | Exposure limit in Luxembourg |
| 0.4 W/m² | Exposure limit in China |
| 0.2 W/m² | Exposure limit in Russia (since 1970), Bulgaria |
| 0.1 W/m² | Exposure limit in Poland, Paris, Hungary |
| 0.1 W/m² | Exposure limit in Italy in sensitive areas |
| 0.095 W/m² | Exposure limit in Switzerland, Italy |
| 0.09 W/m² | ECOLOG 1998 (Germany) Precaution recommendation only |
| 0.001 W/m² | Exposure limit in Austria |

## 4 SMART / ADAPTIVE ANTENNA

The concept of using multiple antennas and innovative signal processing to serve cells more intelligently has existed for many years. In fact, varying degrees of relatively costly smart antenna systems have already been applied in defense systems. Until recent years, cost barriers have prevented their use in commercial systems. The advent of powerful low-cost digital signal processors (DSPs), general-purpose processors (and ASICs), as well as innovative software-based signal-processing techniques (algorithms) have made intelligent antennas practical for cellular communications systems (fig 1). Today, when spectrally efficient solutions are increasingly a business imperative, these systems are providing greater coverage area for each cell site, higher rejection of interference, and substantial capacity improvements [2].

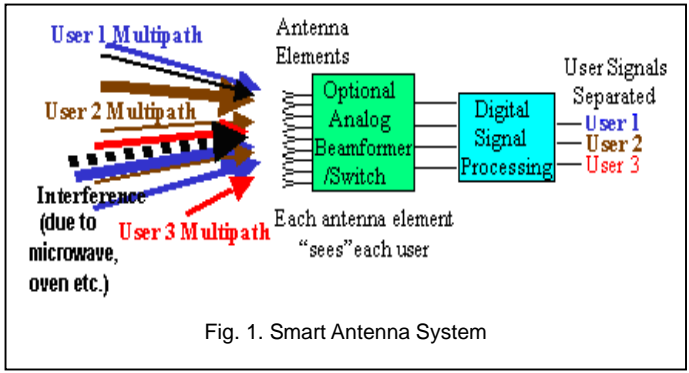

Fig. 1. Smart Antenna System

### 4.1 Merits of Smart Antenna

Smart antennas at base stations can be used to enhance mobile communication systems in several ways:
• Increased BS range
• Less interference within the cell
• Less interference in neighboring cells
• Increased capacity by means of SFIR or SDMA / A-SDMA

## 5 ADAPTIVE SPATIAL DIVISION MULTIPLE ACCESS

A-SDMA is based on the exploitation of the spatial dimension which has so far not been used for parallelism. Using adaptive array antennas at the base station sites, multiple independent beams can be formed with which several users can be served simultaneously on the same radio channel. This approach is an extension of the system where adaptive arrays are used for interference reduction, without exploiting the potential of spatial parallelism. This is illustrated in fig 2. Benefits of A-SDMA are proven and have been demonstrated in [4]. The benefits of adaptive antenna are also proven and can be easily implemented [3]. In essence, the scheme can adapt the frequency allocations to where the most users are located.

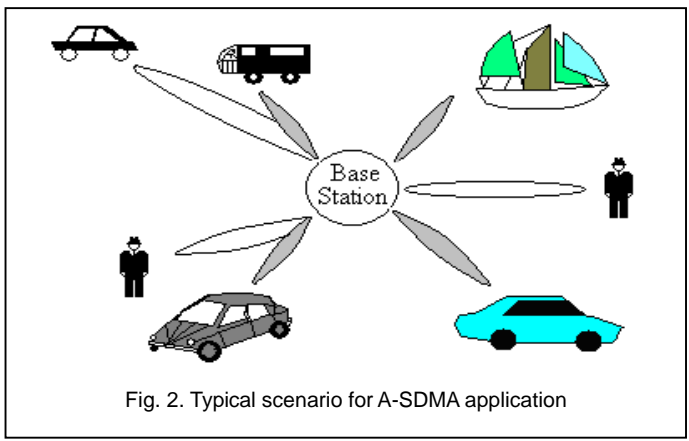

Fig. 2. Typical scenario for A-SDMA application

Because SDMA employs spatially selective transmission, an SDMA base station radiates much less total power than a conventional base station. One result is a reduction in network-wide RF pollution. Another is a reduction in power amplifier size. First, the power is divided among the elements, and then the power to each element is reduced because the energy is being delivered directionally. With a ten-element array, the amplifiers at each element need only transmit one-hundredth the power that would be transmitted from the corresponding single antenna system [5].

## 6 HIERARCHICAL STRUCTURE

Cellular networks are becoming increasingly heterogeneous due to the co-deployment of many disparate infrastructure elements, including micro, pico and femtocells, and distributed antennas. A flexible, accurate and tractable model for a general downlink HCN consisting of K tiers of randomly located BSs, where each tier may differ in terms of average transmit power, supported data rate, and BS density. Assuming 1) a mobile connects to the strongest BS, 2) the target Signal to- Interference-Ratio (SIR) is greater than 0 dB, and 3) received power is subject to Rayleigh fading and path loss. Expressions for the average rate achievable by different mobile users are derived. This model reinforces the usefulness of random spatial models in the analysis and research of cellular networks. This is a baseline tractable HCN model with possible future extensions being the inclusion of antenna sectoring, frequency reuse, power control and interference avoidance/ cancellation [6]. To overcome handoff problem in hierarchical cell structure, efficient use of radio resources is very important. All resources have to be optimally utilized. However, in order to adapt to changes of traffic, it is necessary to consider adaptive radio resource management.

The ability of hierarchical cellular structure (Fig 3) with interlayer reuse to increase the capacity of a mobile communication radio network by applying Total Frequency Hopping (T-FH) and Adaptive Frequency Allocation (AFA) as a strategy to reuse the macro- and micro cell resources without frequency planning in indoor picocells / femtocells have been discussed [7].

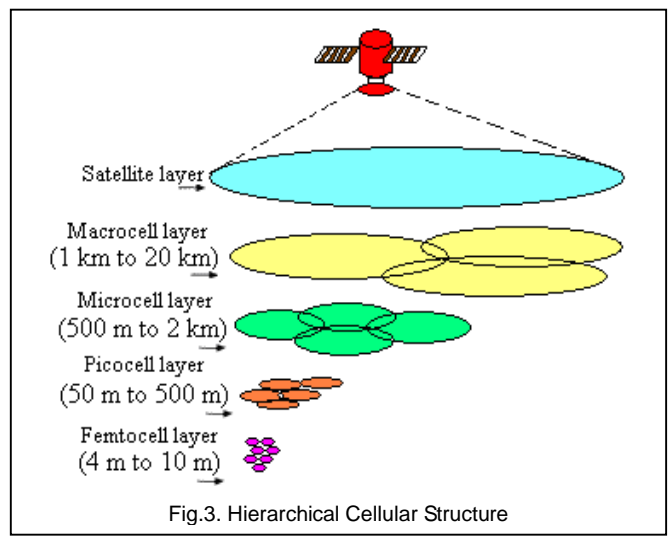

Fig.3. Hierarchical Cellular Structure

### 6.1 Macro Cell

A conventional base station with 20W power and range is about 1 km to 20 km. Macro cell in hierarchical structure takes care of roaming mobiles.

## 6.2 Micro Cell

A conventional base station with 1W to 5W power and range is about 500 m to 2 km. Micro cells and pico cells takes care of slow traffic (pedestrian and in-building subscribers). Micro cells can be classified as the following:

1) Hot Spots: These are service areas with a higher tele-traffic density or areas that are poorly covered. A hot spot is typically isolated and embedded in a cluster of larger cells.

2) Downtown Clustered Micro cells: These occur in a dense, contiguous area that serves pedestrians and mobiles. They are typically found in an "urban maze of" street canyons," with antennas located far below building height.

3) In-Building. 3-D Cells: These serve office buildings and pedestrians (fig 4). This environment is highly clutter dominated, with an extremely high density and relatively slow user motion and a strong concern for the power consumption of the portable units.

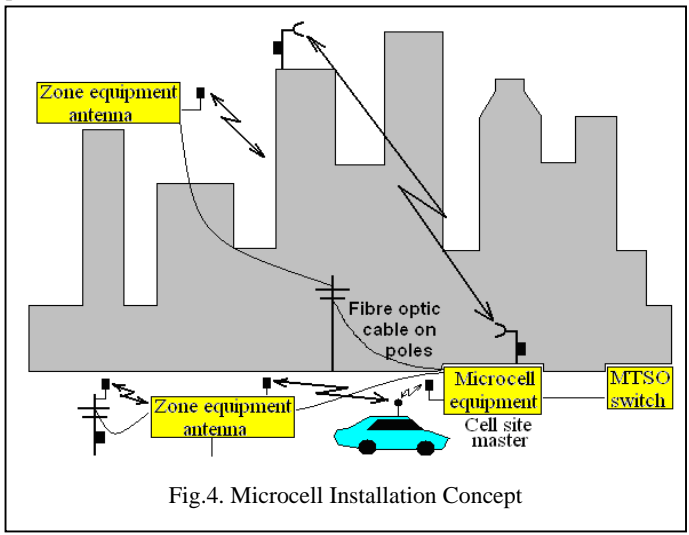

Fig.4. Microcell Installation Concept

## 6.3 Pico Cell

The picocells are small versions of base stations, ranging in size from a laptop computer to a suitcase. Besides plugging coverage holes, picocells are frequently used to add voice and data capacity, something that repeater and distributed antenna cannot do.

Adding capacity in dense area, splitting cells are expensive, time consuming and occasionally impossible in dense urban environment where room for a full size base station often is expensive or unviable. Compact size picocells makes them a good fit for the places needing enhanced capacity, they can get.

Picocells are designed to serve very small area such as part of a building, a street corner, malls, railway station etc. These are used to extend coverage to indoor area where outdoor signals do not reach well or to add network capacity in areas with very dense uses.

## 6.4 Femto Cell

A femtocell is a smaller base station, typically designed for use in home or small business. In telecommunications, a femtocell is a small cellular base station, typically designed for use in a home or small business. It connects to the service provider's network via broadband (such as DSL or cable).

## 7 PROPOSED NETWORK

### 7.1 Evolution Path

For meeting the requirement of RF pollution levels the evolution path is given in fig 4 [4]-

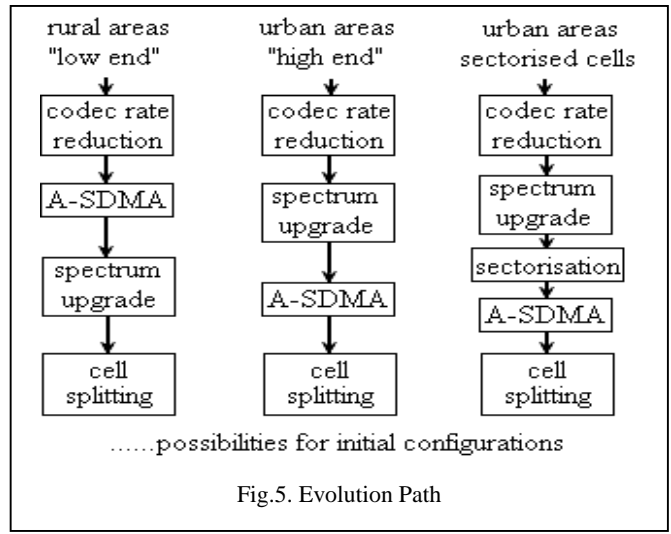

Fig.5. Evolution Path

### 7.2 Network

Proposed network (Fig 4 & 5) is based on simple handoff algorithm discussed in [8] and hierarchical cellular structures with inter-layer reuse in an enhance GSM radio network is suggested [9]. A design of Macro-Micro CDMA Cellular Overlays in the Existing Big Urban Areas is suggested [10]. It is also assumed that the MS is equipped with a Rake receiver capable of performing "maximal ratio combining" of the signals it receives from the transmitting BSs [11]. The following cellular structure will be used for dense urban areas [12]:

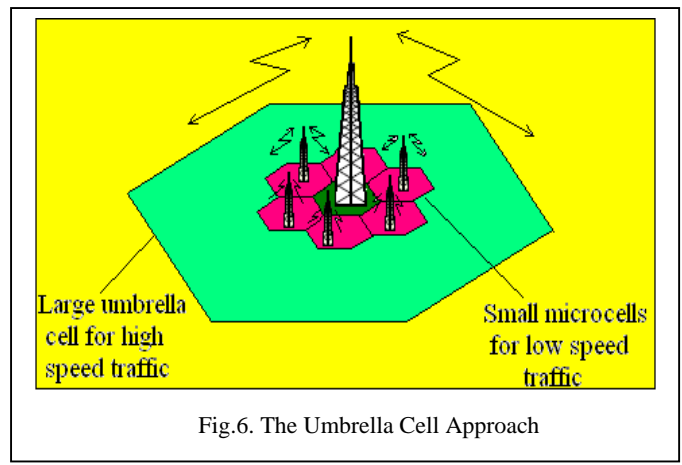

Fig.6. The Umbrella Cell Approach

1) Macro cell will be marked for fast traffic and micro cell will be marked for slow traffic in hierarchical structure. The RF resources will be dynamically allocated between macro and micro cells on the basis of velocity estimation using adaptive array antennas [13].

2) Pico cell will be marked for hotspots. An adaptive frequen-

cy allocation will be applied as strategy to reuse the macro and micro cell resources without frequency planning in indoor picocells [9].

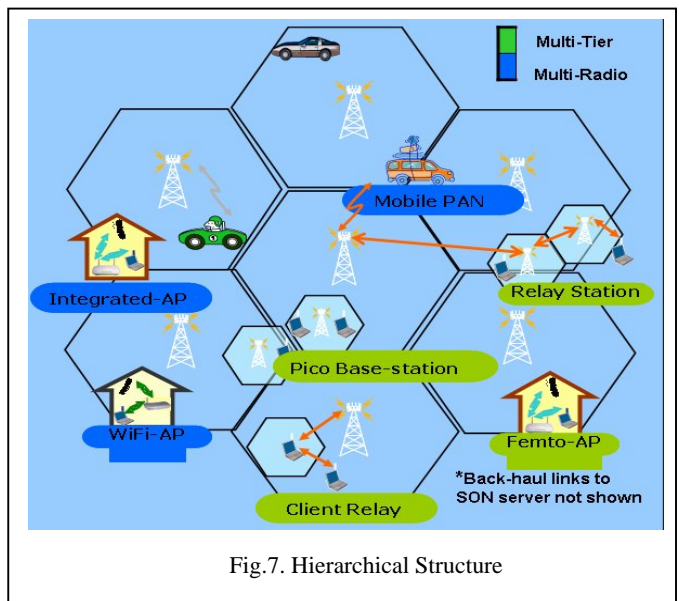

Fig.7. Hierarchical Structure

3) Femto cell will be marked as home base stations- which are data access points installed by home users to get better indoor voice and data coverage. Femtocells enable a reduced transmit power, while maintaining good indoor coverage. Penetration losses insulate the femtocell from surrounding femtocell transmissions. As femtocells serve only around 1-4 users, they can devote a larger portion of their resources (transmit power & bandwidth) to each subscriber. A macro / micro cell, on the other hand, has a larger coverage area (500m-20 km radius), and a larger number of users; providing Quality of Service (QoS) for data users is more difficult. Deploying femtocell will enable more efficient uses of precious power and frequency resources [14].

The above proposed structure will undoubtedly enhance the spectral density with the help of diversity and adaptive approach through rake receiver and adaptive antenna respectively. The induction of pico and femto cell will reuse the RF resources of overlaid macro / micro structures which will enhance spectral density manifold. The simple technologies suggested by William C Y Lee for deployment along city streets, deployment along binding roads, deployment under the ground (subway coverage) and in-building designs etc. will also be considered in proposed hierarchical structure [15].

## 8 CONTRIBUTIONS

I. We present a new optimization model for the hierarchical network design microcell, picocell and femtocell selection. It is a capacity maximization model, for existing 2G / 2.5G system based on proved simple, easily implementable technologies which better reflects the requirements of potential clients.
II. Our basic model is a hybrid network which includes hierarchical structure (umbrella structure), A-SDMA approach along with adaptive frequency allocation and intelligent microcell technologies (such as in-building communication, coverage under the ground, coverage along winding roads, coverage along city streets and simple hand-off technologies).
III. Our model will enhance spectral density and quality of service. RF pollution reduction, cost reduction and power consumption reduction will be the added requirements problem. This is the first 4-tier model that we have seen and it includes macrocell.

## 9 CONCLUSION

Intelligence has been provided in four layers (macro / micro / pico / Femto cell layers) with the help of A-SDMA approach and adaptive frequency allocation along with total frequency hopping techniques. Application of intelligent micro cell, pico cell, Femto cell and adaptive antenna makes this hierarchical structure spectrum efficient. In addition to it power consumption is reduced substantially with the application of adaptive antennas. Low power base stations will result in reduction of interference and R. F. pollution substantially. This has been proved beyond doubt through simulations and practically too by researchers [16] – [18]. Proposed network is not only spectrum efficient but also it will reduce power consumption and R. F. pollution drastically. This will help in achieving our goal of spectral efficiency and reduction of R.F. pollution which is the dire need of hour in India in the prevailing circumstances.